\documentclass[onecolumn,showpacs,preprintnumbers,amsmath,amssymb]{revtex4}
\usepackage{graphicx}% Include figure files
\usepackage{dcolumn}% Align table columns on decimal point
\usepackage{bm}% bold math
\usepackage{color}
\usepackage{amsmath}
\usepackage{xcolor}

\usepackage{caption}
\usepackage{float}
\usepackage{soul}

\usepackage{lipsum}
\usepackage[normalem]{ulem}

\newcommand{\kalpha}{K$_\alpha$\,}
\newcommand{\kbeta}{K$_\beta$\,}

\usepackage{epstopdf}

\begin{document}

\title{X-ray emission from a liquid curtain jet when irradiated by femtosecond laser pulses}

\author{F. Valle Brozas\,$^1$}
\author{D. Papp\,$^2$}
\author{L. M. Escudero \,$^3$}
\author{L. Roso\,$^1$}
\author{A. Peralta Conde\,$^1$}
\affiliation{$^1$ Centro de L\'aseres Pulsados, CLPU, Parque Cient\'ifico, 37185 Villamayor, Salamanca, Spain.\\
$^2$ ELI-ALPS, ELI-HU Non-profit Ltd., 6720 Szeged, Hungary.\\
$^3$ Departamento de Biolog\'ia Celular, Universidad de Sevilla and Instituto de Biomedicina de Sevilla (IBiS), Hospital
Universitario Virgen del Roc\'io/CSIC/Universidad de Sevilla, 41013 Seville, Spain}

\begin{abstract}

Laser-based sources of ionizing radiation have attracted a considerable attention in the last years for their broad potential applications. However the stability and robustness of such sources is still an issue that needs to be addressed. Aiming to solve such problems, we propose a source that uses a liquid jet -rather than a solid- as a target for the production of X-rays. Liquid jets offer always a clean surface for every laser shot which represent a clear advantage over solids. In this work, we present an experimental characterization of the X-ray emission of such targets, and study the efficiency of the process when two temporally delayed pulses are used. According to the obtained results, the X-ray yield is comparable with commonly used targets.
\end{abstract}

\pacs{32.30.Rj, 52.38.Kd, 78.70.En}

\maketitle

\section{Introduction} \label{Introduction}

Cancer is the second leading cause of death in the first world, and is expected to be the leading cause before the end of this century. Despite of the discouraging scenario, over the last decades and triggered by a more detailed knowledge of the interaction between the ionizing radiation and the tumoral tissue, there has been an increase in the average survival of patients diagnosed with such a terrible disease. At the moment, new promising radiotherapy possibilities are attracting large attention of the scientific community. Based on the latest laser technology developments, nowadays is possible to generate ultrashort bunches of ionizing radiation of femtosecond/picosecond time duration with ultrahigh dose-rates 10$^9$ Gy/s, , being 1 gray (Gy) defined as the absorption of one joule of radiation energy per kilogram of matter.  However there are major scientific issues that need to be addressed: can we trigger non-lineal phenomena with this high flux? Is there any advantage to depositing the dose in such ultrashort period of time? Can we deepen our knowledge of the dynamics of irradiated tissues and hence design more efficient radiotherapy treatments and strategies? (see for example \cite{Giulietti16, Rosenwald07, Durante10} and references therein). It is possible to find in literature only few preliminary studies with different cellular lines (mostly in vitro) \cite{Meesat12, Rigaud10, Tillman99, Zeil13, Kraft10, Malka10, Nguyen11, Kong09, Yogo11}. Unfortunately, the results are inconclusive being necessary further research involving the broadest possible collaborations due to the multidisciplinary aspects of the problems. 

So far, different laser-based sources have been proposed but mostly relying on solid targets, and in the direction of achieving challenging performances rather than robustness and user-friendliness \cite{Daido12, Macchi13}. For example, nowadays one can find relatively simple setups capable to produce ultrashort bunches of X-rays and electrons using moderate laser intensities in the range of 10$^{16}$-10$^{17}$\,Wcm$^{-2}$  \cite{Daido12, Macchi13, Tao14, Ivanov11, Hou06, Witte08, Li16, Valle15, Valle16}.  Typically, a femtosecond laser is focalized into a solid target producing at the early stage of the pulse the ionization of the sample. In such conditions, the rest of the pulse interacts predominantly with an expanding plasma coming from the solid target rather than with the target itself. In this interaction, electrons are extracted by the electric field of the laser, accelerated, and reinjected into the plasma bulk. In this process both Bremsstrahlung radiation -by the sudden loss of energy of the re-injected electrons- and the characteristics X-ray emission of the material of the target -by the creation of inner vacant in the atoms of the target- are emitted. Also, electron bunches following the reflection direction of the laser are generated. Both radiations, i.e., electrons and X-rays, inherit the temporal characteristics of the laser pulse. However this scheme presents several difficulties.  Once the laser interacts with the target, this is heavily damage being necessary to move or rotate it to ensure a fresh area for the following laser shot. This limits considerably the reproducibility of the X-ray radiation generated because one can not ensure the same experimental conditions after the target movement. Also the live time, defined as the time of recording events or total time of acquisition, of the possible experiments is reduced by the required target replacement and subsequent realignment. According to this, these sources offer severe limitations for biomedical studies, being necessary a new source generation robust enough to permit systematic measurements and focused on an exquisite control over the radiation characteristics. 

With the intention to offer a solution for the above described problems, we present in this manuscript the generation of laser-based X-rays from a liquid jet curtain \cite{Miaja15, Korn02, Zhavoronkov04, Tompkins98, Papp16}.  The used setup maintains stable the density profile of the liquid shot to shot, allowing a stable X-ray generation without the need of moving and replacing the target. Furthermore, similarly to solid target where the X-rays the emission can be tuned using different materials, for liquids targets this tunability is obtained by solving different salts, for this work we have used KCl and KBr, in an adequate solvent. In the following we will discuss the experimental setup as well as the characterization of the X-ray and electron emission. Finally, we complete our contribution with a brief summary and outlook.

\section{Experimental setup} \label{setup}

The laser used in this work was a commercial Titanium-Sapphire femtosecond system, p-polarized, with a pulse duration of 120\,fs, carrier wavelength 800\,nm, repetition rate 1\,kHz, and energy per pulse up to 7.5\,mJ. The spatial profile was circular with a radius of 0.6\,cm FWHM (Full Width Half Maximum). For the generation of X-rays, just 0.9\,mJ were focused into the target, i.e., into the liquid jet curtain, by  a microscope objective of numerical aperture NA = 0.42. The achieved focal spot measured at low intensity was of the order of $1.5\,\mu$m in the horizontal and $1.2\,\mu$m in the vertical direction being accordingly the expected intensity of the order of $\rm 10^{17}$\,Wcm$^{-2}$. At the conditions for X-ray generation, i.e., using an energy per pulse of 0.9\,mJ, the laser power (P$\approx$7.9\,GW) exceeds the critical power for air (P$\rm _{air}\approx$1.9\,GW), being the focal spot considerably limited by laser filamentation \cite{Chin05}. In these conditions, we estimate a focal spot with a diameter of the order of 100\,$\mu$m and an intensity of the order of $\rm 10^{14}$\,Wcm$^{-2}$. It is noticeable that even at such low intensities X-ray  and electron bunches are produced. The laser was p-polarized, and the angle of incidence into the target was 45 degrees. The X-ray production was detected by an Amptek Silicon drift detector (SDD-132).

\begin{figure}
\includegraphics[width=0.5\columnwidth]{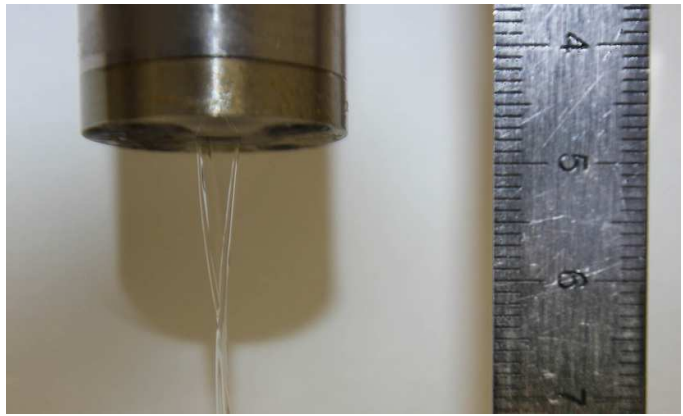}
\caption{\label{fig1} Liquid jet. Dimensions: 4\,mm wide $\times$ $\rm \sim 65\,\mu m$ thick $\times$ 5\,mm long.}
\end{figure}

The liquid jet curtain used as target was generated by a dye circulator connected to a nozzle (Sirah lasertechnik). This kind of circulators are normally used to provide the active medium for CW lasers. However for X-ray production we used rather than the solution of a laser dye in a organic solvent, e.g., ethylene glycol, a mixture of water, glycerin and potassium chloride (KCl) or potassium bromide (KBr). In principle it is possible to use different salts with different concentrations, always that the solution has a density similar to ethylene glycol (1.1132\,g/cm$^3$). Otherwise the liquid jet become unstable and the nozzle can get damage. We achieved that with a mixture of 65\% of glycerol and 35\% of water in weight. Although the solubility of KCl in water is 34.2\,g/100\,g we used just 25\,g/100\,g for precaution. For KBr we just used 28.6\,g/100\,g being the solubility in water 67.8g/100g. The circulator was operating at a stagnation pressure of 10\,bars although it is possible to obtain a stable jet without flickering and/or air bubbles for a wide range of pressures. The dimensions of the jet were $\sim65\,\mu$m thickness times 4\,mm in the widest part (see Fig.\,\ref{fig1}).

\section{Source characterization}

\begin{figure}
\includegraphics[width=0.8\columnwidth]{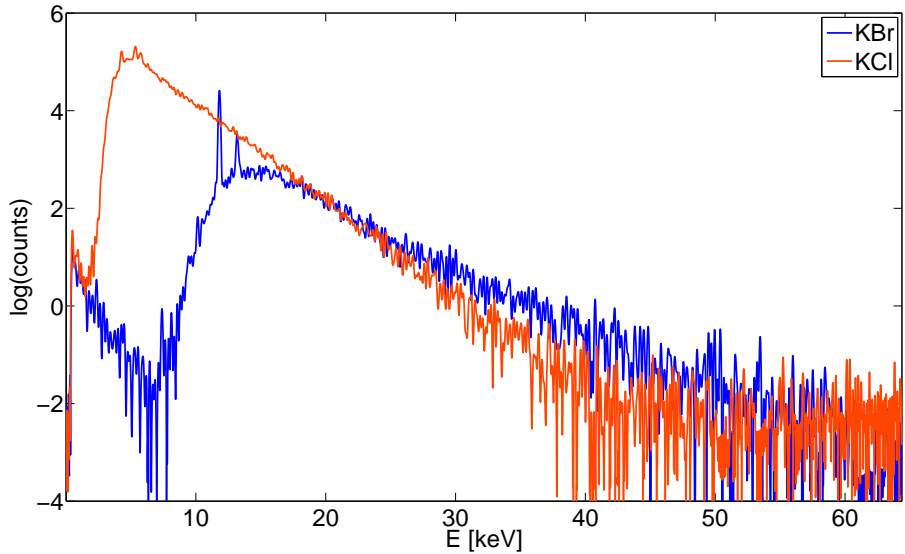}
\caption{\label{fig2} (Color online) Typical X-ray spectrum for a solution of KBr or KCl.}
\end{figure}

Figure\,\ref{fig2} shows a typical X-ray spectrum for a solution of glycerin, water and KBr, and glycerin, water and KCl. It is important to mention for all spectra showed in this work we have apply a Fourier Trasnform Filter (FFT) to avoid the fast oscillations produced by spurious noise. Both spectra are defined by a Bremsstrahlung emission extending to approximately 60 keV produced in the bulk of the liquid jet, and the characteristic X-ray emission of the solute. Concretely, we can see the \kalpha and \kbeta lines of Br at 11.9\,keV and 13.3\,keV. The X-ray emission of Cl and K couldn't be detected (\kalpha=2.6\,keV and \kbeta=2.8\,keV for Cl, and \kalpha=3.3\,keV and \kbeta=3.6\,keV for K) because is absorbed by air and filters \cite{Nist1}. The differences in the spectra for the low energy emission are due to a 1\,cm methacrylate filter used for recording the spectrum of KBr.

\begin{figure}
\includegraphics[width=0.8\columnwidth]{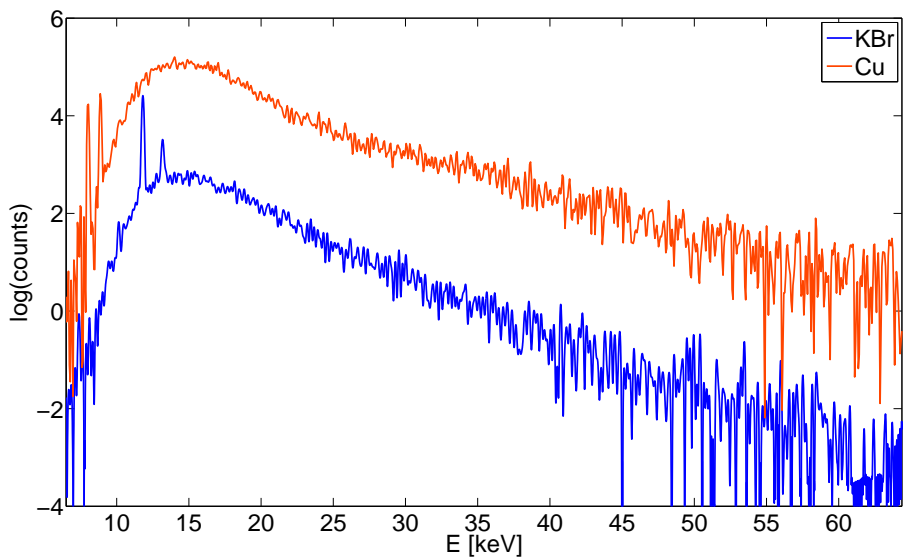}
\caption{\label{fig3} (Color online) Spectra comparison between the emission of a liquid target jet -solution of KBr- and a solid Cu target.}
\end{figure}

To obtain the efficiency of the liquid target, we can compare the \kalpha energy conversion for KBr and for solid Cu targets which have been routinely used so far \cite{Valle15, Valle16}. Figure\,\ref{fig3} shows the spectra used for this comparison. Both measurements were taken in the same conditions: we placed the detector in the laser reflexion direction at a distance of 55 cm from the target. We filtered the radiation with a 1\,cm thickness methacrylate filter to avoid pile up. The mass attenuation coefficient of methacrylate can be adjusted in our region of interest (1-100\,keV) as \cite{Nist1}:

\begin{equation}
\rm \mu/\rho\,(g/cm^2) = 227.16 E^5\,(keV)-1438.4 E^4\,(keV)+4130.2 E^3\,(keV) - 130.63 E^2\,(keV) + 5.54 E\,(keV)+0.116
\end{equation}

To calculate correctly the efficiency of the source, it is necessary to take into account all the elements that reduce the X-ray photon flux prior it reaches the detector: 54\,cm path in air, 1\,cm thickness methacrylate filter, 1.25\,$\mu$m thickness berylium window that protect the detector, and the characteristic quantum efficiency of the detector. Taking all these factors into account, we obtain a total transmission of 1.2$\%$ for 8.0\,keV and 14.6$\%$ for 11.9\,keV.  According to Fig.\,\ref{fig3} we registered 0.0009 counts/s for the \kalpha line of Cu at 8.0\,keV and 0.011 counts/s for the \kalpha line of Br at 11.9\,keV.  If we assume a 4$\pi$ distribution for the \kalpha emission, and consider the effective area of the detector (25\,mm$^2$),  we obtain finally $7.79 \times 10^4$ photons per pulse for Cu \kalpha line and $1.16 \times 10^4$ photons per pulse for Br \kalpha line. This is equivalent to a a efficiency of $1.5 \times 10^{-7}$ for solid Cu and $3.4\times 10^{-8}$ for the liquid solution of KBr. In literature one can find several works reporting the efficiency of different targets in vacuum conditions. For example, for Cu targets (wire and tape) the efficiency range from 10$^{-7}$ to 10$^{-5}$ \cite{Zhanv05, Jiang03, Guo97, Jiang02}, while for molybdenum (Mo) disks one can obtain slightly higher efficiencies of the order of 10$^{-5}$ \cite{Hou06, Li16}. The efficiency of liquid sources in vacuum has been also subject of research. Tompkins \emph{et al} measured an efficiency of 10$^{-8}$ for 200\,$\mu$m water jet with solved copper\,(II) nitrate (Cu(NO$_3$)$_2$) \cite{Tompkins98}, while for a gallium (Ga) jet Zhavoronkov \emph{et al} measured an efficiency of 10$^{-6}$ \cite{Zhavoronkov04}. The efficiency reported in this work for liquid target jets at atmospheric pressure conditions is smaller (10$^{-8}$) than those values. We attribute these differences to the absorption of air and the limitation imposed by filamentation for the minimum achievable focal spot. This leads to a lower intensity and, thus, a lower electron temperature. Despite of this difference, the advantage of working at much simpler experimental conditions and with a target that does not need replacement, compensate by far this yield reduction. 

\begin{figure}
\includegraphics[width=0.8\columnwidth]{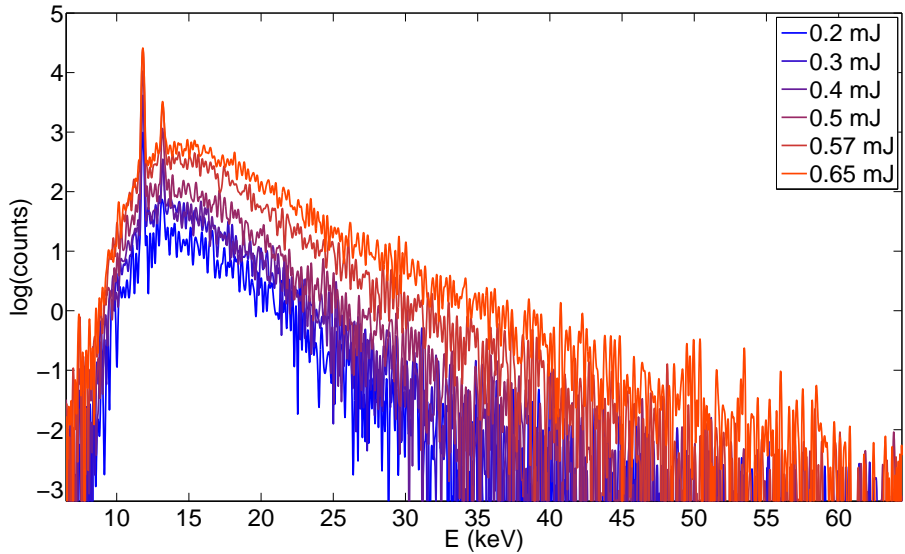}
\caption{\label{Fig4} X-ray spectrum for a KBr solution liquid target as a function of the pulse energy.}
\end{figure}

Aiming to fully characterize the liquid curtain jet source, Fig.\,\ref{Fig4} shows the X-ray yield for KBr solution as a function of the laser pulse energy. Interestingly although the emission yield decrease with the energy, even for low energies per pulse there is an appreciable X-ray generation. Below 0.1\,mJ the X-ray emission was below our setup sensitivity.

\begin{figure}
\includegraphics[width=0.8\columnwidth]{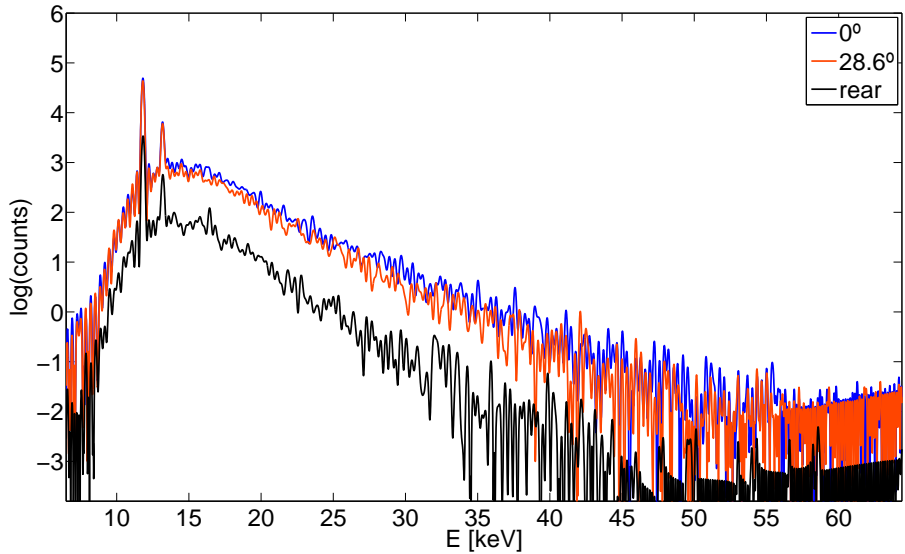}
\caption{\label{Fig5} X-ray spectrum for a KBr solution liquid target as a function of the emission angle.}
\end{figure}

Another interesting aspect to be determined is the directionality of the X-ray emission which can not be properly measured for solid targets. Figure\,\ref{Fig5} shows the X-ray spectrum for different angles of emission being the emission completely isotropic. The slightly smaller yield production for the rear direction, i.e., in the direction of the laser incidence, is produced by the absorption of the target itself. It is interesting to mention that although the \kalpha and \kbeta emission is expected to be isotropic for the Bremsstrahlung this is not so clear. The nature of the Bremsstrahlung emission -this is produced by the sudden loss of energy of accelerated electrons reinjected in the plasma bulk- and the fact that, as we will show later, the accelerated electrons are produced exclusively in the reflection direction of the laser radiation may introduce some sort of anisotropy in the emission. In fact for conventional X-ray tubes, this anisotropy is well established depending on the energy of the electrons \cite{Khan03}. While for few keV electrons the emission is quasi-isotropic, for higher energies the emission is predominant in the forward direction. For these laser-based sources, It is known that the spectra distribution of laser accelerated electrons can be described by two maxwell distribution, corresponding to the electrons that are accelerated by the laser (hot electrons) and the electrons that are accelerated in the plasma expansion (cold electrons) \cite{Gibbon07}. The hot electrons lead to bremsstrahlung distribution that behaves asymptotically like a decreasing exponential. The fitting of this exponential give us a hot electron temperature of 19.6\,keV, which according to \cite{Khan03} is in agreement with an isotropic Bremsstrahlung distribution.

\begin{figure}
\includegraphics[width=0.5\columnwidth]{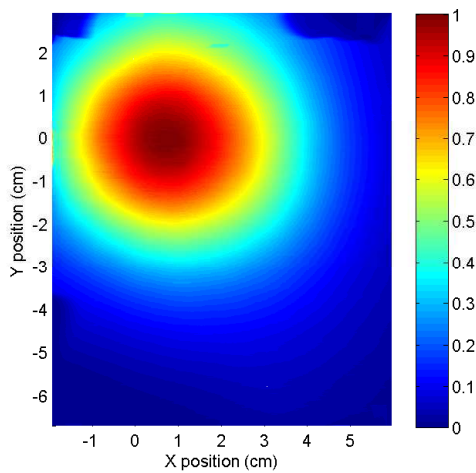}
\caption{\label{Fig6} Dose profile in arbitrary units registered in a gafchromic film for liquid KBr target. }
\end{figure}

Figure\,\ref{Fig6} shows the electron emission direction. These measurements were carried out placing a EBT2 gafchromic film at 5\,cm from the target in the laser reflection direction. These kind of films are very insensitive to the X-ray emission in the range of energies of these work. Thus, we can assure that the signature in the film is exclusively produced by accelerated electrons. As expected, the electron emission is directional corresponding the position (x=0, y=0)  to laser reflection direction.

\section{X-ray emission by double pulses}

The influence of a prepulse in the X-ray generation has been proved in several experiments (see for example \cite{Pelletier97, Lokasani16, Fazeli15, Andreev02, Hatanaka08, Zhu10, Teubner92, Freeman11, Fujioka08}).  The observed enhancement is related with the increased coupling of the stronger second laser pulse with an expanding plasma -created by the first weaker pulse- from the target surface. This expansion takes place in the picosecond time scale, and leads to a volume increase producing a smoother density profile and a reduction in the opacity of the plasma. Since the laser-X-ray conversion efficiency strongly depends of the density gradient of the expanding plasma at a certain delay the laser-plasma coupling is maximize. If the delay between the laser pulses is too small the density profile is steep and there is not any difference in the laser-plasma coupling. On the other hand, if the delay is to large the density drops rapidly and the second laser cannot efficiently heat the plasma electrons with the subsequent reduction in the X-ray emission. For sharp density profiles, i.e., less than the laser wavelength, Brunel absorption is the main interaction laser-plasma mechanism being the energy of the expelled electrons 
\begin{equation}
\rm T_{hot}=8\left(\frac{I}{10^{16}}\lambda^2\right)^{1/3}\,[keV]
\end{equation}
where I is the laser intensity in W/cm$^2$ and $\lambda$ the laser wavelength in $\mu$m \cite{Andreev94}.
However, for smoother density profiles resonance absorption mechanisms prevails being the energy of the expelled electrons slightly higher \cite{Forslund77}
\begin{equation}
\rm T_{hot}\approx14\left(\frac{I}{10^{16}}\lambda^2\right)^{1/3}T_{cold}^{1/3}\,[keV].
\end{equation}

\begin{figure}
\includegraphics[width=\columnwidth]{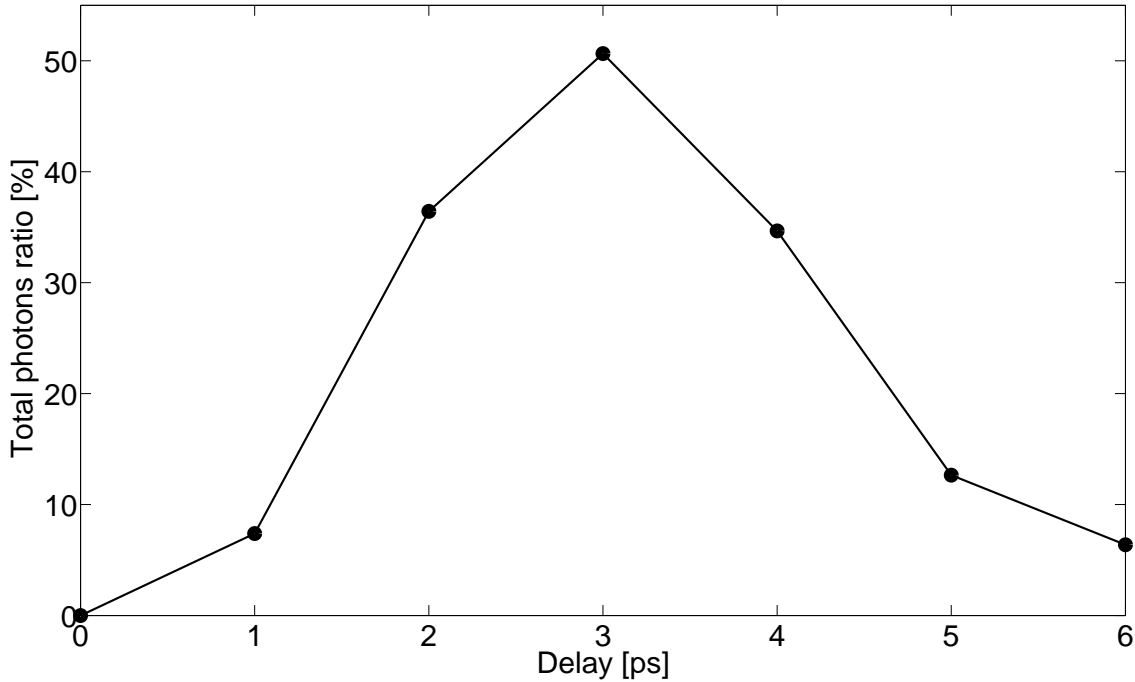}
\caption{\label{Fig7}  X-ray yield increment as a function of the delay between the pulses.}
\end{figure}

To study this phenomenon we slightly modified the setup described above. The original laser radiation was divided using a beam splitter 20:80 into a prepulse and a main pulse. The delay between both pulses was controlled with a precision motorized delay stage with submicron precision. Once delayed one pulse with respect to the other, they were set collinear by a beam combiner, and focalized by the microscope objetive into the same spot. The final energy of the pulses were approximately 0.24\,mJ and 0.10\,mJ.  The different spectra were taken during 5 minutes, at a distance of 55\,cm, in the direction of laser reflection, and filtering de radiation with 1\,cm of methacrylate to avoid pile up. For each delay we recorded the spectrum generated by both pulses independently, and by the combined effect of them. At this low energies, the pre-pulse laser didn't produce any measurable X-ray emission. Figure\,\ref{Fig7} shows the X-ray yield increment with respect to the situation when both pulses are temporally overlapping. We can clearly see for a delay of 3\,ps an increment in the X-ray yield up to 50\%. This result is compatible with the enhancement observed by Hatanaka \emph{et al} in \cite{Hatanaka08} although there are certain differences in the experimental setup. The pulses used in this work are considerable shorter, 120\,fs versus 260\,fs, and the laser was set parallel for both pulses (p-polarized) while in \cite{Hatanaka08} the prepulse is s-polarized and the main pulse p-polarized.

\section{Conclusions} \label{conclusions}

In this work we have shown the potential of liquid jets as a target for X-ray production in air, characterizing their main properties experimentally. Although the X-ray emission efficiency of liquid and solid targets is comparable, the main advantage of liquids is their robustness and stability. Since there is no need to replace the target, as it is the case for solids, it is possible to carry out systematic and/or long-live measurements. These characteristics are mandatory if we want to explore the challenging possibilities that laser-based sources of ionizing radiation may offer for new radiotherapy studies. 

\section{Acknowledgments}
The authors thank J.M. \'Alvarez for fruitful and stimulating discussions during the development of this manuscript. This work has been possible by the support from Ministerio de Educaci\'on Cultura y Deporte (F. Valle Brozas studentship FPU AP2012-3451) and Ministerio de Econom\'ia y Competitividad of Spain  (PALMA project FIS2016-81056-R).


\begin{thebibliography}{7}

\bibitem{Giulietti16}
Laser-Driven Particle Acceleration Towards Radiobiology and Medicine. Editor Antonio Giulietti. Springer International Publishing 2016.

\bibitem{Rosenwald07}
Handbook of Radiotherapy Physics: Theory and Practice. Editors P Mayles, A Nahum, J.C Rosenwald. Taylor \& Francis 2007.

\bibitem{Durante10}
M. Durante, and J.S. Loeffler, Nat. Rev. Clin. Oncol. 7, 37-43, 2010. 

\bibitem{Meesat12}
R. Meesat, H. Belmouaddine, J. F. Allard, C. Tanguay-Renaud, R. Lemay, T. Brastaviceanu, L. Tremblay, B. Paquette, J. R. Wagner, J.-P. Jay-Gerin, M. Lepage, M. A. Huels, and D. Houdea, Proc. Natl. Acad. Sci. USA 109, 38, 2012.

\bibitem{Rigaud10}
O. Rigaud, N.O. Fortunel, P. Vaigot, E. Cadio, M.T. Martin, O. Lundh, J. Faure, C. Rechatin, V. Malka, Y.A. Gauduel, Cell Death Dis. 1, e73, 2010.

\bibitem{Tillman99}
C. Tillman, G. Grafstrom, A. C. Jonsson, B. A. Jonsson, I. Mercer, S. Mattsson, S. E. Strand, and S. Svanberg, Radiology 213, 860-865, 1999.

\bibitem{Zeil13}
K. Zeil, M. Baumann, E. Beyreuther, T. Burris-Mog, T. E. Cowan, W. Enghardt, L. Karsch, S. D. Kraft, L. Laschinsky, J. Metzkes, D. Naumburger, M. Oppelt, C. Richter, R. Sauerbrey, M. Sch\"urer, U. Schramm, J. Pawelke, Appl. Phys. B, 110, 4, 2013.

\bibitem{Kraft10}
S. D. Kraft, C. Richter, K. Zeil, M. Baumann, E. Beyreuther, S. Bock, M. Bussmann, T. E. Cowan, Y. Dammene, W. Enghardt, U. Heibig, L. Karsch, T. Kluge, L. Laschinsky, E. Lessmann, J. Metzkes, D. Naumburger, R. Sauerbrey, M. Schürer, M. Sobiella, J. Woithe, U. Schramm, and J. Pawelke,  New J. Phys., vol. 12, 2010.

\bibitem{Malka10}
V. Malka, J. Faure, and Y. A. Gauduel, Rev. Mutat. Res., vol. 704, no. 1–3, pp. 142–151, 2010.

\bibitem{Nguyen11}
J. Nguyen, Y. Ma, T. Luo, R. G. Bristow, D. A. Jaffray, and Q.-B. Lu, Proc. Natl. Acad. Sci. U. S. A. 108, 11778, 2011.

\bibitem{Kong09}
X. Kong, S. K. Mohanty, J. Stephens, J. T. Heale, V. Gomez-Godinez, L. Z. Shi, J. S. Kim, K. Yokomori, M. W. Berns, Nucleic Acids Res 37, 9, 2009.

\bibitem{Yogo11}
A. Yogo, T. Maeda, T. Hori, H. Sakaki, K. Ogura, M. Nishiuchi, A. Sagisaka, H. Kiriyama, H. Okada, S. Kanazawa, T. Shimomura, Y. Nakai, M. Tanoue, F. Sasao, P. R. Bolton, M. Murakami, T. Nomura, S. Kawanishi, and K. Kondo, Appl. Phys. Lett. 98, 053701, 2011.

\bibitem{Daido12}
H. Daido, M. Nishiuchi, and A. S. Pirozhkov, Prog. Rep. Phys. 75, 2012.

\bibitem{Macchi13}
A. Macchi, M. Borghesi, M. Passoni, Rev. Mod. Phys. 85, 2013. 

\bibitem{Tao14}	
B. B. Zhang, S. S. Sun, D. R. Sun, and Y. Tao, Rev. Sci. Instrum. 85, 096110, 2014.

\bibitem{Ivanov11}
K.A. Ivanov, D.S. Uryupina, R.V. Volkov, A.P. Shkurinov, I.A. Ozheredov, A.A. Paskhalov, N.V. Eremin, and A.B. Savel\'eva, Nuc. Inst. Meth. Phys. Res. A 653, 58–61, 2011.

\bibitem{Witte08}
H. Witte, M. Silies, T. Haarlammert, J. H\"uve, J. Kutzner, H. Zacharias, Appl. Phys. B 90, 11-14, 2008.

\bibitem{Valle15}
F. Valle Brozas, A. V. Carpentier, C. Salgado, J. I. Apinaniz, M. Rico, M. S\'anchez Albaneda, J. M. \'Alvarez, A. Peralta Conde, and L. Roso, Rev. Esp. Fis. 29-3, 2015. 

\bibitem{Valle16}
F. Valle Brozas, A. Crego, L. Roso, and A. Peralta Conde, Appl. Phys. B 122:220, 2016.

\bibitem{Li16}
M. Li, K. Huang, L. Chen, W. Yan, M. Tao, J. Zhao, Y. Ma, Y. Li, and J. Zhang, Radiat. Phys. Chem. 2016. 

\bibitem{Hou06}
B. Hou, J. Nees, A. Mordovanakis, M. Wilcox, G. Mourou, L.M. Chen, J.-C. Kieffer, C.C. Chamberlain, and A. Krol, Appl. Phys. B 83, 1, 81-85, 2006.

\bibitem{Korn02}
G. Korn, A. Thoss, H. Stiel, U. Vogt, M. Richardson, T. Elsaesser, and M. Faubel, Opt. Lett. 27, 866, 2002.

\bibitem{Zhavoronkov04}
N. Zhavoronkov, Y. Gritsai, G. Korn and T. Elsaesser, App. Phys. B 79, 663-667, 2004.

\bibitem{Tompkins98}
R. J. Tompkins, I. P. Mercer, M. Fettweis, C. J. Barnett, D. R. Klug, Lord G. Porter, I. Clark, S. Jackson, P. Matousek, A. W. Parker, and M. Towrie, Rev. Scient. Instr. 69, 3113, 1998.

\bibitem{Miaja15}
L. Miaja-Avila, G. C. O\'Neil, J. Uhlig, C. L. Cromer, M. L. Dowell, R. Jimenez, A. S. Hoover, K. L. Silverman, and J. N. Ullom, Struct. Dyn. 2, 024301, 2015.

\bibitem{Papp16}
D. Papp, R. Polanek, Z. Lecz, L. Volpe, A. Peralta Conde, and A. A. Andreev, IEEE Trans. Plasma Science, 44, 10, 2016.

\bibitem{Chin05}
S. L. Chin, S. A. Hosseini, W. Liu, Q. Luo, F. Th\'eberge, N. Ak\"ozbek, A. Becker, V. P. Kandidov, O. G. Kosareva, and H. Schroeder, Can. J. Phys. 83, 863-905, 2005.

\bibitem{Nist1}
NIST. X-ray Mass attenuation coefficients. 
https://www.nist.gov/pml/x-ray-mass-attenuation-coefficients

\bibitem{Zhanv05}
N. Zhavoronkov, Y. Gritsai, M. Bargheer, M. Woerner, and T. Elsaesser, Appl. Phys. Lett. 86, 244107, 2005.

\bibitem{Jiang03}
Y. Jiang, T. Lee, and C.G. Rose-Petruck, J. Opt. Soc. Am. B 20, 229–237, 2003.

\bibitem{Guo97}
T. Guo, C. Rose-Petruck, R. Jimenez, F. Raksi, J. A. Squier, B. C. Walker, K. R. Wilsom , and C. P. J. Barty, SPIE 3157, 84, 1997.

\bibitem{Jiang02}
Y. Jiang, T. Lee, W. Li, G. Ketwaroo, Chr. G. Rose-Petruck, Opt. Lett. 27, 963, 2002.

\bibitem{Khan03}
Faiz M. Khan, The Physics of Radiation Therapy, Lippicott Williams \& Wilkins, 2003

\bibitem{Gibbon07}
P. Gibbon, Short Pulse Laser Interaction with Matter, Imperial College Press, 2007

\bibitem{Pelletier97}
J. F. Pelletier, M. Chaker, and J. C. Kieffer, J. Appl. Phys.,81, 5980, 1997

\bibitem{Lokasani16}
R. Lokasani,, G. Arai, Y. Kondo, H. Hara, T.-H. Dinh, T. Ejima, T. Hatano, W. Jiang, T. Makimura, B. Li, P. Dunne, G. O'Sullivan, T. Higashiguchi, and J. Limpouch, Appl. Phys. Lett. 109, 19, 2016.

\bibitem{Fazeli15}
R. Fazeli and M. H. Mahdieh, Phys. of Plasmas, 22, 11, 2015.

\bibitem{Andreev02}
A. Andreev, J. Limpouch, A. B. Iskakov, and H. Nakano, Phys. Rev. E, 65, 026403, 2002.

\bibitem{Hatanaka08}
K. Hatanaka, H. Ono, and H. Fukumura, Appl. Phys. Lett. 93, 0641013, 2008.

\bibitem{Zhu10}
P. Zhu, Z. Zhang, L. Chen, J. Zheng, R. Li, W. Wang, J. Li, X. Wang, J. Cao, D. Qian, Z. Sheng, and J. Zhang, Appl. Phys. Lett. 97, 211501, 2010.

\bibitem{Teubner92}
U. Teubner, G. K\"uhnle, and F. P. Sch\"afer, Appl. Phys. B 54, 493, 1992.

\bibitem{Freeman11}
J. R. Freeman, S. S. Harilal, and A. Hassanein, J. Appl. Phys. 110, 083303, 20011.

\bibitem{Fujioka08}
S. Fujioka, M. Shimomura, Y. Shimada, S. Maeda, H. Sakaguchi, Y. Nakai, T. Aota, H. Nishimura, N. Ozaki, A. Sunahara, K. Nishihara, N. Miyanaga, Y. Izawa, and K. Mima, Appl. Phys. Lett. 92, 241502, 2008.

\bibitem{Andreev94}
A. A. Andreev, J. Limpouch, and A. N. Semakhin, Bull. Russ. Acad. Sci., 58:1056–1063, 1994.

\bibitem{Forslund77}
D. W. Forslund, J. M. Kindel, and K. Lee, Phys. Rev. Lett., 39:284–288, 1977.

\end{thebibliography}
\end{document}